\documentclass{kapproc} % Computer Modern font calls
\let\footnote\savefootnote
\let\footnotetext\savefootnotetext 
\let\footnoterule\savefootnoterule 
\kluwerbib

\RequirePackage{graphicx}%
\RequirePackage{epsf}%

\begin{document}

\articletitle{Simulating the Hot X-ray Emitting Gas in Elliptical Galaxies}

\author{Daisuke Kawata and Brad K.\ Gibson}

\affil{Centre for Astrophysics \& Supercomputing,
Swinburne University, Australia}
\email{dkawata,bgibson@astro.swin.edu.au}

\chaptitlerunninghead{Cosmological TreeSPH Simulations of Ellipical Galaxies}

\begin{abstract}
We study the chemo-dynamical evolution of elliptical galaxies and
their hot X-ray emitting gas using high-resolution cosmological
simulations. Our Tree N-body/SPH code includes a self-consistent treatment of
radiative cooling, star
formation, supernovae feedback, and chemical enrichment. We present a
series of ${\rm \Lambda}$CDM cosmological simulations which trace the
spatial and temporal evolution of heavy element abundance patterns in both the
stellar and gas components of galaxies. X-ray spectra of the hot gas
are constructed via the use of the {\tt vmekal} plasma model, and analysed
using XSPEC with the XMM EPN response function. 
Simulation end-products are quantitatively compared with the 
observational data in both the X-ray and optical regime.
We find that radiative cooling is important to interpret the
observed X-ray luminosity, temperature, and metallicity of the interstellar
medium of elliptical galaxies.
However, this cooled gas also leads to excessive
star formation at low redshift, and therefore results in underlying galactic
stellar populations which are too blue with respect to observations.
\end{abstract}

\section{Introduction}

The hot X-ray emitting gas of elliptical galaxies represents an
important interface between galaxies and the intergalactic medium (perhaps 
even the {\it primary} interface).  The X-ray halos of ellipticals carry
with them two fundamental mysteries:
\begin{itemize}
\item their X-ray luminosities are lower than that expected from an
extrapolation of the cluster X-ray luminosity-temperature 
(${\rm L_X}-{\rm T_X}$) relation (e.g. Matsushita et~al. 2000).
\item their X-ray metallicities are lower than that of the mean
stellar iron abundance (the so-called ``iron discrepancy'' -
e.g. Arimoto et~al. 1997 - a ``discrepancy'' in the sense that the halo gas
metallicity was expected to exceed that of the stars, since it should bear
the pollution of the enrichment from earlier generation of stars - 
enrichment byproducts that were not locked up into subsequent stellar
generations).\footnote{X-ray iron abundances remain a controversial
issue (c.f. Buote \& Fabian 1997), although the iron discrepancy appears to
hold based upon recent high-resolution XMM RGS 
observations (Xu et~al. 2002; Sakelliou et~al. 2002).}
\end{itemize}
Conversely, the optical properties of ellipticals appear less contentious!
The Colour-Magnitude Relation (CMR)
and Fundamental Plane provide strong constraints for any elliptical
galaxy formation paradigm.  We present here our preliminary work aimed 
ultimately at the construction of successful self-consistent
optical $+$ X-ray cosmological chemodynamical simulations of elliptical
galaxies.

\begin{figure}[t]
\plotone{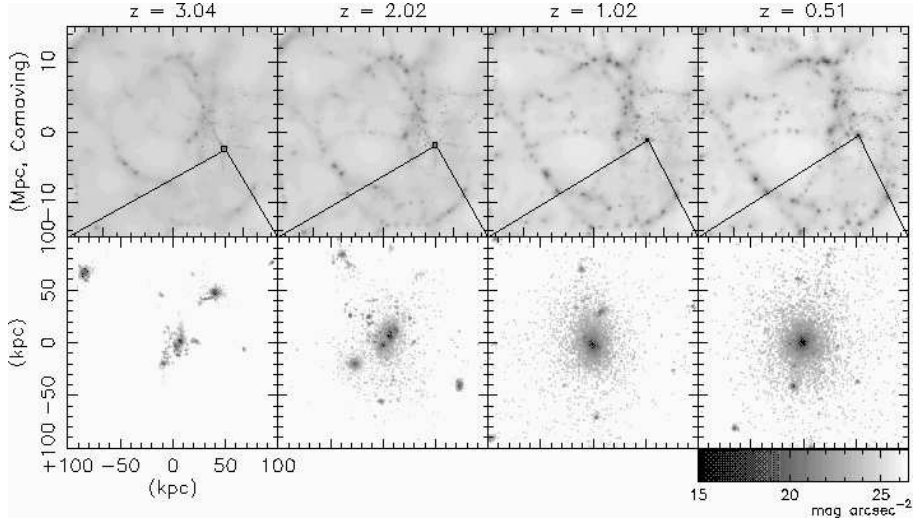}
\caption{Dark matter density map of a portion of the 43~Mpc (comoving) 
simulation volume
({\it upper panels}), and predicted $I$-band image of the target galaxy
({\it lower panels}), over the redshift range $z$=3.0 to $z$=0.5.
}
\label{evol-fig}
\end{figure}

\section{Methods}

In our simulations, the dynamics of collisionless dark
matter and stars is calculated using a gravitational Tree N-body code,
and the gas component is modeled using Smoothed Particle Hydrodynamics
(SPH).  We calculate radiative cooling, star formation, chemical enrichment, 
and supernovae (SNe) feedback, self-consistently, and take into account both 
Type~Ia and Type~II SNe.  
%We assume that SNe
%feedback is released as a combination of thermal
%($E_{\rm th}$) and kinetic ($E_{\rm kin}$) 
%energy, and their ratio is given by the parameter
%$f_v=E_{\rm kin}/(E_{\rm th}+E_{\rm kin})$;
%$f_v$ acts essentially as a SNe feedback ``control knob'' (Kawata 2001).
%The role of feedback ``strength'' will be commented upon later.
We assume that SNe feedback is released as thermal energy.
Details of the code are presented in Kawata (2001)
and Kawata \& Gibson (2003, in prep).

We have carried out a series of high-resolution
simulations within the adopted standard $\Lambda$CDM cosmology
($\Omega_0$=0.3, $\Lambda_0$=0.7, 
$\Omega_{\rm b}$=0.019$h^{-2}$, $h$=0.7, and $\sigma_8$=0.9). 
Gas dynamics and star formation are included only within the relevant
high-resolution region
($\sim$12~Mpc at $z$=0); the surrounding low-resolution region ($\sim$43~Mpc)
contributes to the high-resolution region only through gravity.
The mass of individual gas particles in the high-resolution region 
was $5.9\times10^7$~${\rm M}_\odot$.  We next identified an appropriate
elliptical galaxy analog in the high-resolution region, which acts as
the focus for this preliminary study.
The total virial mass of this target galaxy is
$2\times10^{13}$~${\rm M_\odot}$, similar in size to that of
NGC~4472, a bright
elliptical galaxy in the Virgo Cluster.  The target galaxy is relatively
isolated, with only a few low-mass satellites remaining at $z$=0.

Figure \ref{evol-fig} shows the morphological evolution of 
dark matter in the simulation volume, and the evolution of the stellar
component in a 200~kpc region centred on the target galaxy.
The galaxy forms through conventional hierarchical clustering
between redshifts $z$=3 and $z$=1; the morphology has not changed dramatically
since $z$=1.  Three different radiative cooling and SNe feedback models
were considered:
Model~A is an adiabatic model (i.e. no cooling);
Model~B includes cooling and weak feedback;
% Model~C mimics Model~B, but incorporates stronger feedback.
Model~C mimics Model~B, but incorporates stronger feedback
(100 times larger thermal energy per supernova).

For all the models, we examine both the resulting X-ray {\it and}
optical properties, comparing them quantitatively with observation.
The gas particles in our simulations carry with them knowledge of the 
density, temperature, and abundances of various heavy elements in their
immediate vicinity.
Using the XSPEC {\tt vmekal} plasma model, we derive the X-ray spectra
for each gas particle, and synthesize them within the assumed aperture
%(R$\sim$35~kpc). 
(R$\sim$20~kpc). 
We next generate ``fake'' spectra with the response function of XMM EPN
detector, assuming an exposure time (40~ks) and target galaxy distance
(17~Mpc).  Finally, our XPSEC fitting provides the X-ray weighted
temperatures and abundances of various elements.  Conversely, the
simulated star particles each carry their own age and metallicity ``tag'',
which enables us to
generate an optical-to-near infrared spectral energy distribution for
the target galaxy, when combined with our population synthesis code
adopting simple stellar populations of Kodama \& Arimoto (1997).

\begin{figure}[t]
\plottwo{lxt.ps}{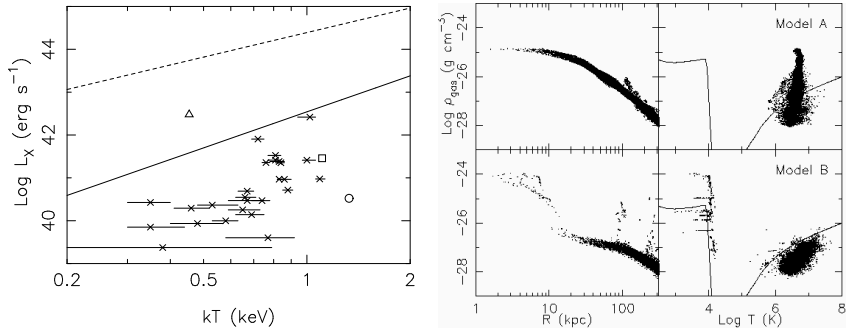}
\caption{ {\it Left panel}: Comparison of the simulated and observed
(crosses with error bars) ${\rm L_X-T_X}$ relations.
The triangle/circle/square indicates the predictions of Model~A/B/C.
The solid line indicates the extrapolation of the cluster ${\rm L_X-T_X}$
relation (Edge \& Stewart 1991), while the
dashed line represents the relation obtained from the adiabatic
simulation of Muanwong et~al. (2001).
{\it Right panel}: Density vs radius (left) and density vs
temperature (right) distributions of gas particles 
for Model~A (upper) and Model~B (lower). The sold curves separate the
region where the cooling time is shorter (upper region) and longer
than the Hubble time.
}
\label{lxt-fig}
\end{figure}

\begin{figure}[t]
\plottwo{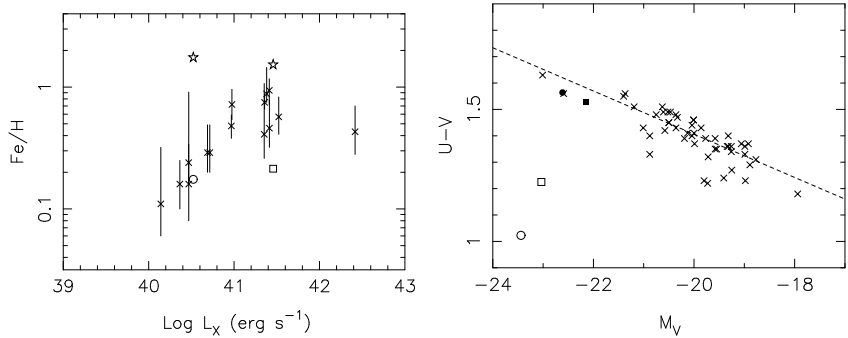}{cmr.ps}
\caption{ {\it Left panel}: Comparison of the simulated and
observed (crosses with error bars) ${\rm [Fe/H]_X-L_X}$ relations.
The open circle/square shows the predictions of Model~B/C;
open stars show the predicted mean metallicity of the stellar component
for Models~B and C.
{\it Right panel}:  Comparison of the simulated CMRs 
(open circle/square for Model~B/C) and
that of the Coma cluster ellipticals (crosses).
The dashed line shows the CMR fitted to the Coma Cluster galaxies.
The solid circle/square demonstrates the colour and magnitude
for Model~B/C when the contribution from young stars (age $<$8~Gyr) 
is ignored. 
}
\label{lxfecmr-fig}
\end{figure}

\section{Results and Conclusion} 

The left panel of Figure~\ref{lxt-fig} shows 
the predicted ${\rm L_X-T_X}$ relation for the three models
at $z$=0; crosses with error bars represent the observational data
from Matsushita et~al. (2000).
The adiabatic model (Model~A) appears incompatible with
the data due to its excessive luminosity and low temperature.
The inclusion of radiative cooling leads to lower 
luminosities and higher temperatures -  as a result,
models with cooling (Models B and C) are (roughly) consistent with
the ${\rm L_X-T_X}$ relation of the observed elliptical galaxies.
These conclusions are consistent with the analysis of Muanwong et~al.
(2001).
The right panel of Figure~\ref{lxt-fig} shows the effect of cooling
more clearly. In the gas density versus temperature diagrams,
the region above than the line corresponds to a parameter space in
which the cooling time is shorter 
than the Hubble time. Cooling ensures the gas within this region
is cold (ie. non X-ray emitting), and of
low density and high temperature, ensuring that ultimately radiative
cooling drives the observed ${\rm L_X-T_X}$ relation.

The left panel of Figure~\ref{lxfecmr-fig} compares the X-ray weighted 
iron abundance of our simulations with the observational data of 
Matsushita et~al. (2000).  As the adiabatic model (by construction) does
not form any stars (not having any cooling!), we show
the results only for Models~B and C.  Both these models show 
lower gas-phase iron abundance, 
compared to their stellar abundance, consistent with the low 
iron abundances observed in the X-ray emitting gas of ellipticals.
We find that a large fraction of iron ejected from stars is locked
into future generation of stars.  Stars preferentially
enrich the gas in the central
region, where cooling is efficient (right panel of Figure~\ref{lxt-fig}). 
The enriched gas can then
cool easily and be incorporated into future generations of stars.
Consequently, the hot gaseous halo has not been enriched efficiently, 
leading to a lower X-ray weighted iron abundance.

In summary, our radiative cooling models explain the two 
X-ray ``mysteries'' alluded to in Section~1.  Having said that, any
successful scenario must also explain the optical properties of the
underlying stellar component.
To this end, we examined the position of our simulated target galaxy
in the observed Coma cluster CMR (Bower et~al. 1992).
We can see immediately that the colours of the resulting stellar
components of both Models~B and C are inconsistent with the data (being
too blue).  This inconsistency can be traced to 
an excessive population of young and intermediate age stars ($<$8~Gyr)
which form from successively cooled gas,
regardless of the strength of SNe feedback.
If the contribution of these young stars was ignored, the resulting
colours would match the observed CMR.  Therefore, one exotic (if 
somewhat {\it ad hoc}) solution is to ``hide'' these younger stars within
a bottom-heavy initial mass function (IMF) such that they cannot be
seen today even if they did exist
(e.g. Fabian et~al. 1982; Mathews \& Brighenti 1999).
Another (more plausible) possibility is that extra heating
sources, such as intermittent AGN activity, suppress star formation
at low redshift.  Before suggesting this is the true solution though,
we must re-examine the predicted X-ray properties of the simulation
end-products after introducing these additional heating sources; we
will be pursuing this comparison in a future paper.

%\section{Conclusion}

Our cosmological chemodynamical code makes it possible to undertake
quantitative comparisons between numerical simulations
and observational data in both the X-ray and optical regime
with minimal assumptions.
We find that radiative cooling is required to explain the
observed X-ray luminosity, temperature, and metallicity of 
elliptical galaxies.  Unfortunately, the resulting cooled gas also leads
to unavoidable overproduction of young and intermediate age stellar
populations, at odds with the observational constraints.
Although a bottom-heavy IMF is one solution for this problem,
extra heating by intermittent AGN activity seems to be more
plausible (e.g. Brighenti \& Mathews 2002);
recent observations are consistent with this latter picture
(e.g. Churazov et~al. 2001).

\section*{Acknowledgements}
We acknowledge the support of the Australian Research Council through the 
Large Research Grant Program (A0010517) and Swinburne University through the
Research Development Grants Scheme.


\begin{chapthebibliography}{1}
\bibitem[Arimoto et al.\ (1997)]{amior97}
 Arimoto, N., Matsushita, K., Ishimaru, Y., et~al.,
 1997, ApJ, 477, 128
\bibitem[Bower, Lucey, \& Ellis (1992)]{ble92}
 Bower, R.G., Lucey, J..R., \& Ellis, R.S., 1992, MNRAS, 254, 589
\bibitem[Brighenti \& Mathews (2002)]{bm02}
 Brighenti, F., \& Mathews, W.G., 2002, ApJL, 574, L11
\bibitem[Buote \& Fabian (1998)]{bg98}
 Buote, D.A., Fabian, A.C., 1998, MNRAS, 296, 977
\bibitem[Churazov et al.\ (2001)]{cbkbf01}
 Churazov E., Br\"uggen, M., Kaiser, C.R., et~al.,
 2001, ApJ, 554, 261
\bibitem[Edge \& Stewart (1991)]{es91}
 Edge, A.C., \& Stewart, G.C., 1991, MNRAS, 252, 414
\bibitem[Faian, Nulsen, \& Canizares (1982)]{fnc82}
 Fabian, A.C., Nulsen, P.E.J., \& Canizares, C.R., 1982, MNRAS, 201, 933
\bibitem[Kawata (2001)]{dk01}
 Kawata, D., 2001, ApJ, 558, 598
\bibitem[Kodama \& Arimoto (1997)]{ka97}
 Kodama, T., \& Arimoto, N., 1997, A\&A, 320, 41 
\bibitem[Matsushita, Ohashi, \& Makishima (2000)]{mom00}
 Matsushita, K., Ohashi, T., \& Makishima, K., 2000, PASJ, 52, 685
\bibitem[Mathews \& Brighenti (1999)]{mb99}
 Mathews, W.G., \& Brighenti, F., 1999, ApJ, 526, 114
\bibitem[Muanwong et al.\ (2001)]{mtkpc01}
 Muanwong, O., Thomas, P.A., Kay, S.T., et~al.,
 2001, ApJL, 552, L27
\bibitem[Sakelliou et al.\ (2002)]{spt02}
% Sakelliou, I.\ et al., A\&A in press,
% (astro-ph/0206249)
 Sakelliou, I.\ et al., 2002, A\&A, 391, 903
\bibitem[Xu et al.\ (2002)]{xkp02}
% Xu, H.\ et al., ApJ in press
% (astro-ph/0110013)
 Xu, H.\ et al., 2002, ApJ, 579, 600

\end{chapthebibliography}
\end{document}